\documentclass[aps,pre,amsfonts,floatfix,superscriptaddress,longbibliography,twocolumn,footinbib]{revtex4-2}
\usepackage[utf8]{inputenc}
\usepackage{amsmath}    
\usepackage{amssymb}
\usepackage{mathbbol}
\usepackage{graphicx}   
\usepackage{verbatim}   
\usepackage{subfigure}  
\usepackage{hyperref} 
\usepackage{scalerel}
\usepackage{cancel}
\usepackage{soul}
\usepackage[normalem]{ulem}
\usepackage[dvipsnames]{xcolor}
\usepackage{siunitx}
\usepackage{physics}
\usepackage{enumerate} 
\usepackage{empheq}

\begin{document}
\title{From pink to brown: Interaction-driven noise color shift in the Aubry-Andr{\'e} Model}
\author{M. Jim{\'e}nez-Valdez}
\affiliation{Institute of Physics, Benem{\'e}rita Universidad Aut{\'o}noma de Puebla, 72570 Puebla, Pue., Mexico}
\author{S. A. Montes-Camacho}
\affiliation{School of Physics, Universidad Industrial de Santander, 680002 Bucaramanga, Colombia}
\author{E. J. Torres-Herrera}
\affiliation{Institute of Physics, Benem{\'e}rita Universidad Aut{\'o}noma de Puebla, 72570 Puebla, Pue., Mexico}

\begin{abstract}
We revisit the metal-insulator transition in a one-dimensional lattice subjected to an on-site quasiperiodic potential, modeled using spin-1/2 particles with tunable nearest-neighbor interactions. To characterize the transition, we examine the power-spectrum, defined as the squared modulus of the Fourier transform of consecutive level spacings. In the absence of interactions, the power-spectrum fails to capture clear signatures of the transition, consistently exhibiting brown-noise-like behavior across regimes. In contrast, the interacting case reveals a marked change: the metallic phase displays spectral fluctuations consistent with pink noise, while the insulating phase is characterized by brown noise. These results are supported by analyses of the level spacing distribution and the level number variance. Additionally, we demonstrate that both statistics and the structure of energy eigenstates provide reliable indicators of the transition, in both interacting and non-interacting scenarios.
\end{abstract}

\maketitle

\section{Introduction}

The study of metal-insulator transitions in low-dimensional systems has long been central to understanding quantum localization phenomena~\cite{Anderson1958,Mott1968,Evers2008,Abrahams2010}. A paradigmatic model in this context is the Aubry-André (AA) model, which describes a one-dimensional (1D) tight-binding chain subject to an incommensurate on-site potential. This model exhibits a sharp transition from extended to localized single-particle states as the potential strength is varied, making it a rare example of a disorder-driven metal-insulator transition occurring in one dimension without the need for randomness~\cite{Harper1955,Aubry1980}.

While the non-interacting AA model has been extensively analyzed and even realized experimentally, for instance with microwaves~\cite{Kuhl1998}, quasi-periodic Bose-Einstein condensates~\cite{Roati2008} or superconducting qubits~\cite{Li2023}, the inclusion of interactions introduces significant complexity and leads to richer physical behavior. In particular, interactions can give rise to many-body localization (MBL), a phenomenon where quantum systems fail to thermalize, retaining memory of their initial conditions even at long times~\cite{Basko2006,Oganesyan2007,Nandkishore2015,Torres2015,Abanin2019}. In the interacting AA model, the competition between a quasiperiodic potential and interactions leads to various regimes, namely, extended, localized, and critical, whose identification requires precise diagnostic tools~\cite{Luitz2015}.

Spectral statistics provide a powerful approach for probing such transitions. Measures like the level spacing distribution and level number variance have been widely used to distinguish between ergodic and localized phases~\cite{Oganesyan2007,Atas2013,Serbyn2016}.
However, these standard indicators may not capture the noise-like structure present in spectral fluctuations.

$1/f$-like noise is ubiquitous in nature, like in music~\cite{Voss1978}, astronomy \cite{Press1978}, solid-state quantum information~\cite{Paladino2014} or in physical, chemical, and biological contexts~\cite{Weissman1988,West1990}.

In this work, we revisit the metal-insulator transition in both the non-interacting and interacting Aubry-André models by studying the power-spectrum of energy-level fluctuations. The power-spectrum, defined as the squared modulus of the Fourier transform of the sequence of consecutive level spacings, provides valuable information about the correlations within the spectrum. In this context, the role played by time in conventional power-spectrum is taken by the index labeling the position of each level in the ordered sequence. Different regimes of spectral statistics manifest as distinct types of so-called colored noise: for instance, chaotic systems typically display $1/f$ (pink) noise, while integrable or localized systems are characterized by $1/f^2$ (brown) noise~\cite{Relano2002,Gomez2005,Faleiro2006,Relano2008,Pachon2018}. The power-spectrum approach has been applied to level statistics of microwave graphs~\cite{Bialous2016} and, more recently, in studies related to MBL~\cite{Corps2020,Corps2021} and to diagnose the existence of extended non-ergodic states in the so-called $\beta$-ensembles~\cite{Das2022}.

Our results show that in the absence of interactions, the power-spectrum fails to detect the metal-insulator transition, exhibiting brown-noise-like behavior throughout. In contrast, when interactions are introduced, the transition becomes apparent: in the metallic regime, the power-spectrum displays features consistent with pink noise, while in the insulating regime, it resembles brown noise. These findings are supported by complementary analyses of the level spacing distribution and level number variance. Additionally, we demonstrate that the statistical properties and structural features of energy eigenstates can also distinguish the phases, in both non-interacting and interacting scenarios.

By emphasizing the role of interactions in making the power-spectrum sensitive to the transition, our study contributes to the broader understanding of spectral diagnostics and their relevance to quantum localization in quasiperiodic systems.

The models studied in this work are described in Sec. ~\ref{sec:models}. The quantities employed are defined in Sec.~\ref{sec:quantities}. We present the results in Sec.~\ref{sec:results} and finalize with Conclusions in Sec.~\ref{sec:condisc}. 


\section{Models}\label{sec:models}
We consider a system of spin-1/2 particles arranged on a one-dimensional lattice of $L$ sites with an on-site quasi-periodic potential, described by the Hamiltonian
\begin{equation}
H\!=\!\! \sum_{j=1}^{L} (S_j^xS_{j+1}^x\! +\! S_j^yS_{j+1}^y\! +\! \Delta S_j^zS_{j+1}^z)\!+\!\sum_{j=1}^{L} h_jS^z_j.
\label{eq:HAA}
\end{equation}
Where $S^{x,y,z}$ are spin-$1/2$ operators, $j$ is an integer number which labels the sites, $L$ is the number of sites. $\Delta$ is the anisotropic parameter. When $\Delta=0$ the model corresponds to a non-interacting system, meanwhile when $\Delta\neq0$ the model includes Ising-like interactions. $h_{j}=h\cos(2\pi j\beta+\phi)$ is the quasi-periodic Harper potential, with $h$ as the potential strength, $\beta$ an irrational number which we take as the inverse golden mean, $\beta=(1-\sqrt{5})/2$. Closed boundary conditions are imposed, $S^{x,y,z}_{L+1}=S^{x,y,z}_1$, just to avoid possible boundary effects. 

\begin{figure}[ht!]
    \centering
    \includegraphics[scale=0.4]{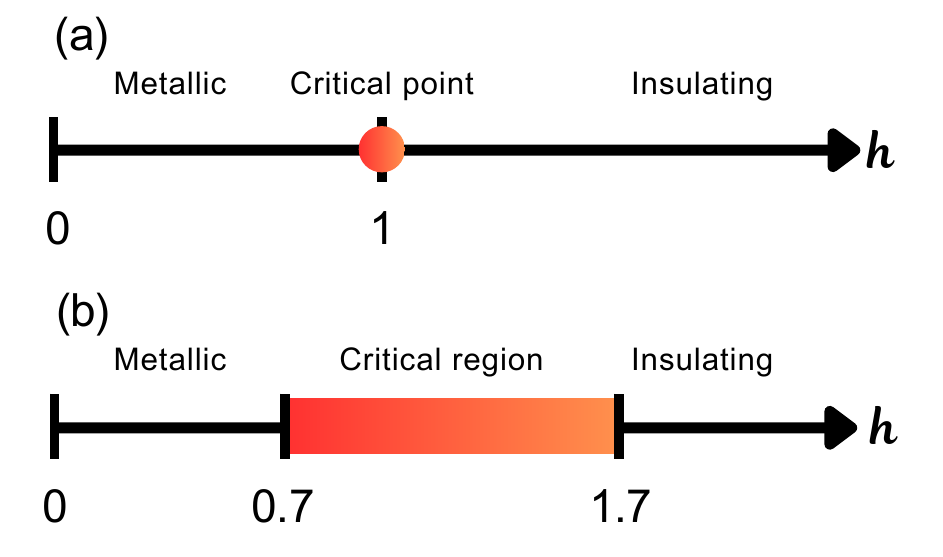}
    \caption{Illustrative scheme of the metal-insulator transition in both models, non-interacting (a) with a single critical point and interacting (b) with a critical region.
    }
    \label{fig:Dia_AAM}
\end{figure}

It is well established that the non-interacting model depicts a metal-insulator transition at $h=1$. The interacting model apparently undergoes two distinct transitions: the first, from a metallic phase to a critical phase occurring at $h\approx 0.7$, while the second transition, from the critical phase to a many-body localized phase at $h\approx 1.7$~\cite{Xu2019}. These two transitions are illustrated in Fig.~\ref{fig:Dia_AAM}.
Hamiltonian given in Eq.~\eqref{eq:HAA} is rotationally invariant about the $z$-direction, so it commutes with the total magnetization in that direction, ${\cal{S}}^{z}=\sum_{k=1}^{L}S_{k}^{z}$. We focus on the symmetry sector with ${\cal{S}}^{z}=0$, where the dimension of the corresponding subspace of the Hilbert space is ${\cal{N}}=L!/(L/2)!^2$, the largest one which leads to better statistics.

\section{Quantities}\label{sec:quantities}
This section is devoted to introduce the main quantities that will be analyzed in this work. All quantities in this section are defined in terms of energy eigenvalues (energy levels) and eigenstates (energy states) of the system represented by Hamiltonian~\eqref{eq:HAA}, that is, those involved in the stationary version of the Schr{\"o}dinger equation
\begin{equation}
H|\psi_\alpha\rangle=E_\alpha|\psi_\alpha\rangle,\quad\alpha=1,\,2,\,\dots,\, {\cal{N}}.
    \label{eq:Schro}
\end{equation}
Where $\{E_\alpha\}$ and $\{|\psi_\alpha\rangle\}$ are the energy levels and energy states respectively. They will be obtained by exact diagonalization.

\subsection{Power-spectrum}
The first diagnostic that we will employ to characterize the metal-insulator transition is the so-called power-spectrum of the statistics $\delta_n$, which is defined through
\begin{equation}
    \delta_n= \sum_{\alpha=1}^n(s_\alpha-\overline{s} )= \epsilon_{n+1}\!-\!\epsilon_1\!-\!n,\,\,\, n=1\,,2,\,\dots,\, {\cal{N}}-1.
    \label{eq:deltas}
\end{equation}
Where $s_\alpha=(E_\alpha-E_{\alpha-1})/\overline{s}$ are the consecutive spacings between adjacent energy levels, $\overline{s}$ corresponds to the mean level spacing and $\left\{\epsilon_\alpha=E_\alpha/\overline{s}\right\}$ are the \emph{unfolded} energy levels. The unfolding procedure is needed to remove peculiarities of the system and reveal universal features of the energy spectrum at the scale of the mean level spacing.

Provided with $\delta_n$, its Fourier transform is computed
\begin{equation}
\tilde{\delta}_k= \dfrac{1}{\sqrt{\cal{N}}} \sum_{n=0}^{{\cal{N}}-1} \delta_n \exp\left(-2\pi i \dfrac{k n}{{\cal{N}}}\right).
\label{eq:Fouriertransform}
\end{equation}
Followed by its power-spectrum, defined through 
\begin{equation}\label{eq:psgen}
 P_k= \left|\tilde{\delta}_k\right|^2. 
\end{equation}
Analytical expressions of $P_k$ are already known for Poisson-like spectra and for the Gaussian orthogonal ensemble (GOE) from random matrix theory (RMT)~\cite{Faleiro2004},
\begin{equation}\label{eq:PkPoiss}
  \left\langle P_k \right\rangle_\text{Poisson} = 
        \dfrac{1}{4\sin^2{(\frac{\pi k}{\mathcal{D}}})}.   
\end{equation}
\begin{equation}\label{eq:PkGOE}
\begin{split}
\left\langle P_k \right\rangle_\text{GOE}&=\dfrac{1}{4\sin^2{(\frac{\pi k}{\mathcal{D}}})} -\dfrac{1}{12}+ \\ 
&+ \dfrac{\mathcal{N}^2}{4\pi^2}\left( \dfrac{K(\frac{k}{\mathcal{N}})-1}{k^2}
        +\dfrac{K(1-\frac{k}{\mathcal{N}})-1}{(\mathcal{N}-k)^2}\right).
\end{split}        
\end{equation}
Where $K(x) = [1-2x + x \ln(1+2 x)] \Theta (1- x) + \{-1 + x \ln [ (2x+1)/(2x -1) ] \} \Theta(x-1)$ is the two-level form factor for GOE, in terms of the Heaviside $\Theta$ function. $K$ is responsible for the correlation hole appearing in the dynamical behavior of the survival probability of physical systems~\cite{Torres2017ptrs,Torres2018} and trivially manifested, as a particular case, in the well known so-called spectral form factor~\cite{Talib2025}. In Eqs.~\eqref{eq:PkPoiss} and~\eqref{eq:PkGOE} the symbol $\langle\dots\rangle$ stands for an average over disorder realizations in the context of RMT.

Both expressions lead respectively to asymptotic power-law decays 
\begin{equation}\label{eq:PkaGOE}
\left\langle P_k \right\rangle_\text{Poisson}= \dfrac{{\cal{N}}^2}{4\pi^2}\dfrac{1}{k^2} \quad\text{and}\quad\left\langle P_k \right\rangle_\text{GOE}= \dfrac{1}{k},
\end{equation}
characteristic of brown and pink noise respectively. One could expect a similar distinction between metallic and insulating phases.

\subsection{Level spacing distribution}
Energy level repulsion is considered as a main fingerprint of a regime with extended states, but this has not been proven rigorously to be a one to one correspondence. After de-symmetrization of the system and unfolding of the spectrum~\cite{Guhr1998,Santos2009,Gubin2012}, short-range correlations in the form of level repulsion can be detected, for instance, by computing the nearest-neighbor level spacing distribution $P(s)$, where $s$ are the consecutive spacings like in Eq.~\eqref{eq:deltas}. For GOE matrices, as well as for chaotic Hamiltonians belonging to the same symmetry class, the level spacing distribution coincides with Wigner's surmise,
\begin{equation}
 {P_\text{GOE}(s)= \frac{\pi }{2} s \exp \left(- \frac{\pi }{4} s^2 \right)}.
 \label{eq:PsWD}
\end{equation}
Where linear repulsion is apparent in the limit $s\to 0$, where $P(s)\to s$~\cite{McDonald1979,Casati1980,Bohigas1984}.

In integrable or localized systems, usually the levels are not prohibited from crossing and the distribution typically coincides with that related to the consecutive events of a Poisson process~\cite{Berry1977,Sorensen1991,Hofstetter1993,Oganesyan2007,Sorathia2012,Serbyn2016,Torres2019,Torres2020}, 
\begin{equation}
   P_\text{Poisson}(s)=\exp(-s). 
   \label{eq:PsPoi}
\end{equation}
Specifically, the exponential distribution shows level clustering at $s=0$. Deviations from Poisson statistics in integrable systems exist, a prominent example is the quantum harmonic oscillator whose energy levels are equally spaced and show the strongest degree of repulsion, even more than the spectrum of any classical RMT ensemble. For other examples, see for instance~\cite{Casati1985} and in particular for localized systems~\cite{Yang2024}. Although some of those exceptions could only be part of subsets with measure zero~\cite{Scaramazza2016} or non-generic features of a specific integrable or localized model~\cite{Feingold1985,Seligman1986}.

\subsection{Level number variance}
Information about long-range correlations among the energy levels, a signature of spectral rigidity, is manifested in so-called level number variance given by 
\begin{equation}
\Sigma^2 (\ell)  \equiv (\overline{N(\ell ,\epsilon))^2}  - \left( \overline{N(\ell ,\epsilon)}\right)^2,    
\end{equation}
where $N(\ell ,\epsilon)$ stands for the number of unfolded energy levels $\epsilon$ in an energy window of width $\ell$, and the bar denotes the average over different intervals for a fixed $\ell$. 
For matrices from GOE, it is given by
\begin{equation}
    \Sigma_\text{GOE}^2 (\ell) = \frac{2}{\pi^2} \left[ \ln(2 \pi \ell) + \gamma + 1 -\frac{\pi^2}{8} \right], 
    \label{eq:LNVGOE}
\end{equation}
where $\gamma = 0.57721\ldots $ is known as the Euler-Mascheroni constant~\cite{Mehta2004}. Equation~\eqref{eq:LNVGOE} represents strong spectral rigidity. For finite-size chaotic physical models, $\Sigma_\text{GOE}^2 (\ell)$ for $\ell > E_\text{Th}$ does not follow the behavior predicted by Eq.~\eqref{eq:LNVGOE}, but instead a behavior proportional to $(\ell/E_\text{Th})^{1/2}$ for a particle in a one-dimensional system~\cite{Altshuler1986}, where $E_\text{Th}$ is so-called Thouless energy which is inversely proportional to the time for a single electron to diffuse through the sample~\cite{Thouless1977}. The concept of Thouless energy has been extended to the realm of many-body quantum systems and in general it is defined as the energy scale beyond which level statistics is not described by RMT, see for instance Refs.~\cite{Bertrand2016,Vsuntajs2020,Corps2020,Schiulaz2019}.

Uncorrelated levels with Poisson-like statistics lead to 
\begin{equation}
 \Sigma^2_\text{Poisson} (\ell) = \ell.   
\end{equation} 
That is, a linear behavior with respect to $\ell$.

\subsection{Participation ratio}
Given a quantum state $|\psi_\alpha \rangle = \sum_n  C_\alpha^n |\phi_n \rangle$ with $C_\alpha^n\equiv \langle\phi_n|\psi_\alpha \rangle$, the participation ratio, $\text{PR}_\alpha$, is defined as
\begin{equation}
\text{PR}_\alpha = \frac{1}{\sum_{n=1}^{\cal{N}} |C^n_\alpha |^4}.
\label{Eq:PR}
\end{equation}
It measures how much extended (delocalized) a state is in the basis vectors $|\phi_n \rangle$. An state is completely extended if it visits the whole Hilbert space without any preference, which gives $\text{PR}_\alpha = {\cal{N}}$. Typical examples of extended and chaotic states are those of full random matrices, like the ones of GOE matrices, which behave as vectors with uncorrelated random components with Gaussian distribution and are extended with $\text{PR}_\alpha^\text{GOE}=({\cal{N}}+2)/3$~\cite{Ulla196465}.

\subsection{Porter-Thomas distribution}
For extended states it is expected that the probabilities $ |C^{n}_{\alpha}|^2$ follow the Porter-Thomas distribution~\cite{Porter1965,Brody1981RMP,Zelevinsky1996},
\begin{equation}\label{eq:PTD}
P_{PT} (|C_{\alpha}^{n}|^2) = \left(\frac{{\cal{N}}}{2\pi |C_\alpha^n|^2}\right)^{1/2}\exp[\left(-\frac{{\cal{N}}}{2}|C_\alpha^n|^2\right)].
\end{equation}
The Porter-Thomas distribution is a result from nuclear physics, in particular for the description of resonance width fluctuations~\cite{Porter1956}. In the context of RMT and quantum chaos, it has become an additional tool to determine if a state is chaotic. The components of localized states deviate from the Porter-Thomas distribution~\cite{Torres2017}.

\section{Results}\label{sec:results}
In this section, we present results obtained from the numerical computation of the quantities introduced and described in the previous section. For all quantities we fix $L=16$ leading to ${\cal{N}}=12870$, a sufficiently large number to ensure good statistical accuracy and reliable results.

\subsection{Energy level statistics}
\begin{figure}[ht!]
    \centering
    \includegraphics[scale=0.3]{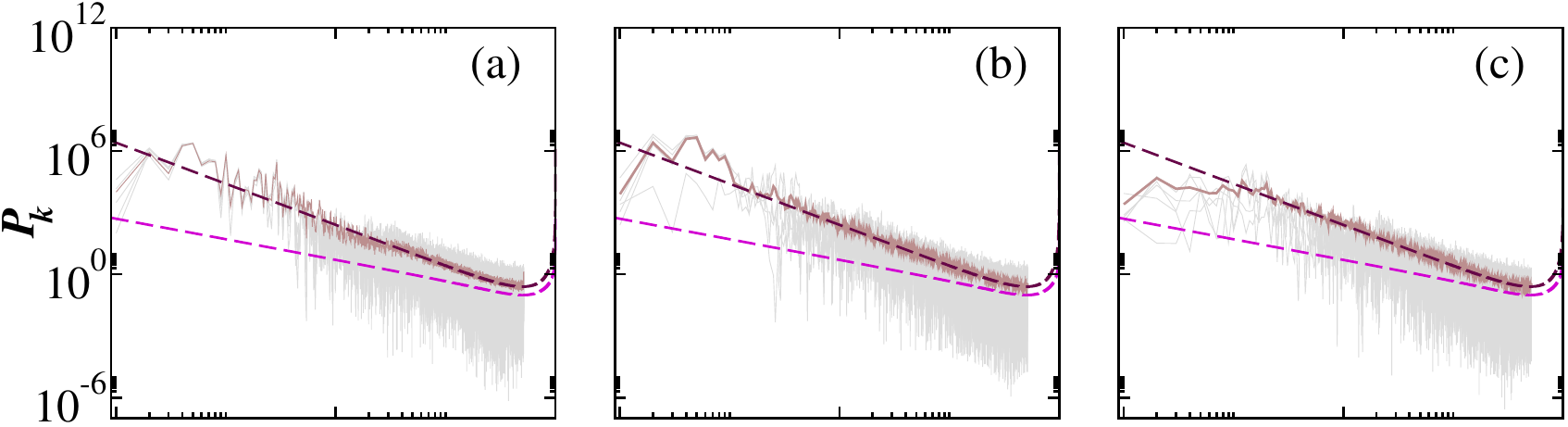}
    \includegraphics[scale=0.3]{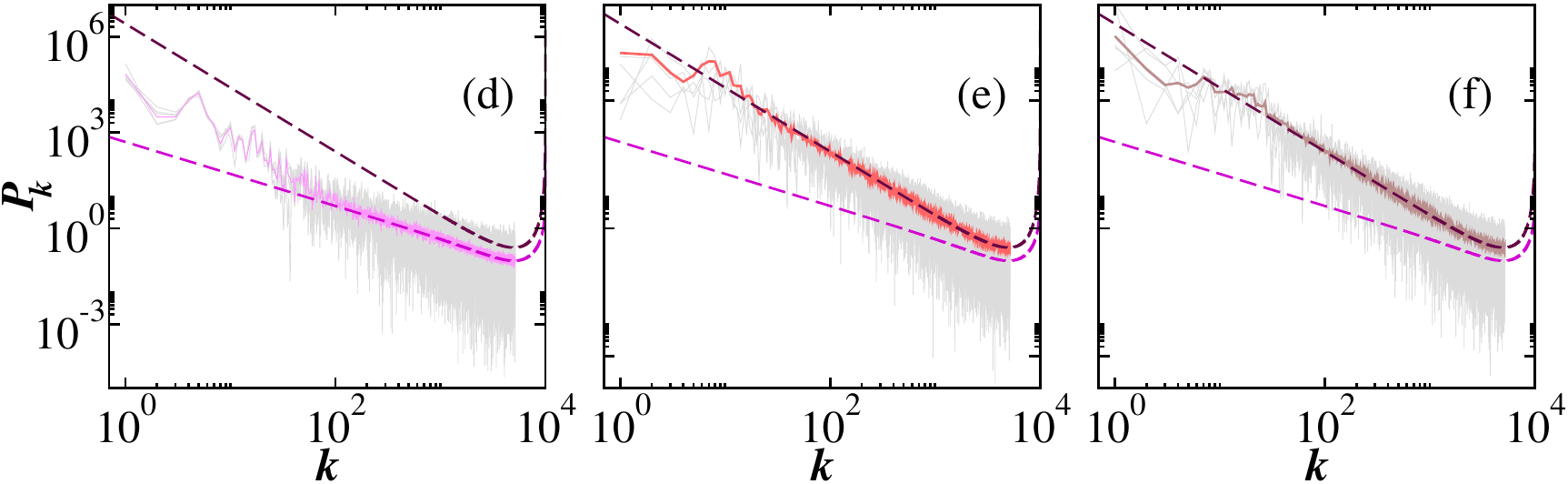}
    \caption{Power-spectrum for the non-interacting (top) and interacting (bottom) AA model. (a) $h=0.2$, (b) $h=1.0$, (c) $h=3.0$, (d) $h=0.2$, (e) $h=1.2$, (f) $h=3.0$. Pink dashed curves are given by Eq.~\eqref{eq:PkGOE}, while brown dashed curves by Eq.~\eqref{eq:PkPoiss}. Gray curves show individual samples and solid curves are averages over $10^3$ samples. 
    }
    \label{fig:PSNIAAM}
\end{figure}

Figure~\ref{fig:PSNIAAM} shows results for the power-spectrum of energy levels from the non-interacting AA model (top panels) and the interacting AA model (bottom panels). Three different strengths of the Harper potential are shown, each one is representative of the different phases or regimes where the system is, (a, d) extended, (b, e) critical and (c, f) localized.

We include curves for some individual samples (gray color), just to show sample-to-sample fluctuations of the power-spectrum. After averaging over $10^3$ samples the power-law decay is apparent. The results are compared with the analytical expressions given by Eqs.~\eqref{eq:PkPoiss} and~\eqref{eq:PkGOE} for Poisson- (brown dashed curves) and GOE-like (pink dashed curves) statistics respectively.

\begin{figure}[ht!]
    \centering
    \includegraphics[scale=0.3]{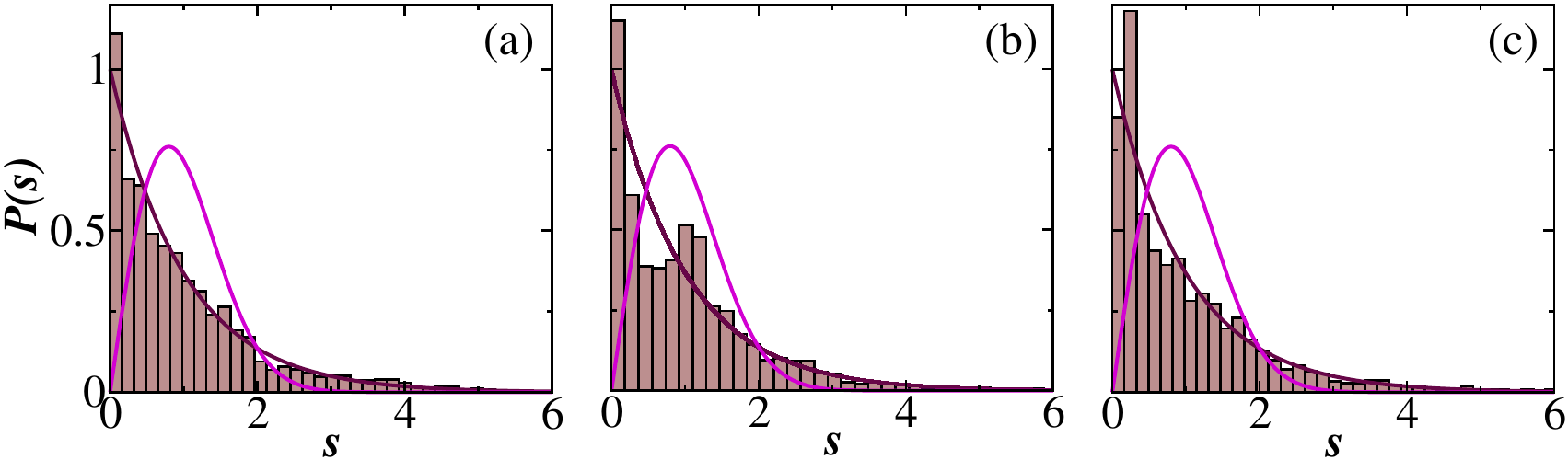}
    \includegraphics[scale=0.3]{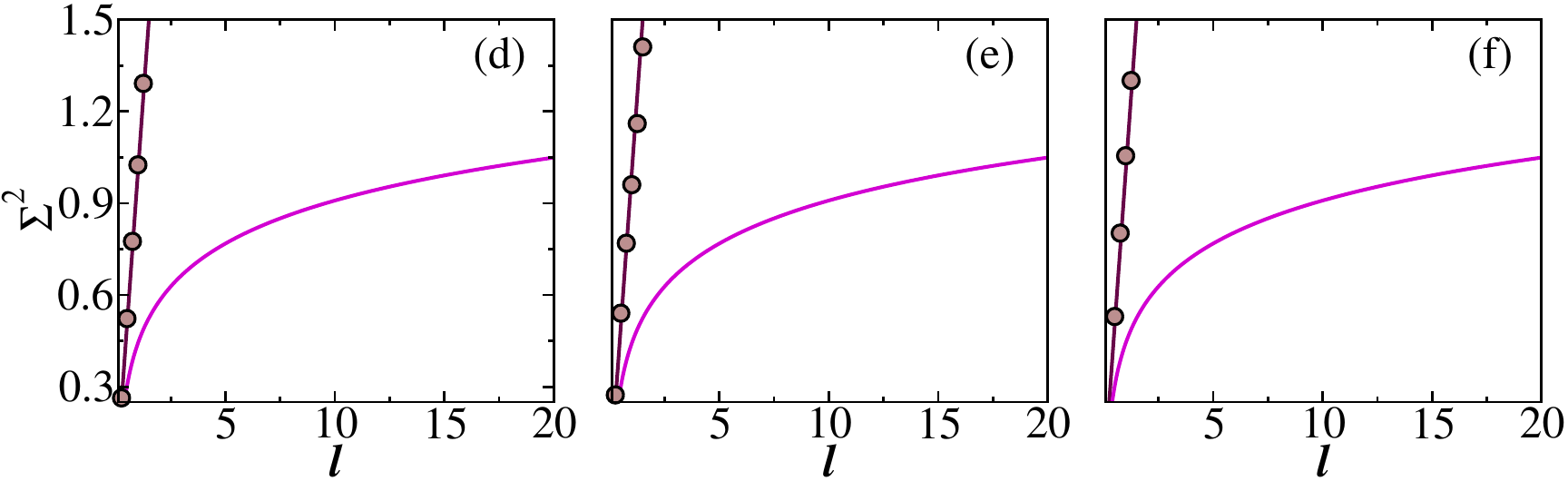}
    \caption{Distribution of consecutive level spacings (top) and level number variance (bottom) for the non-interacting AA model. Pink solid curves show the Wigner surmise (top) and $\Sigma^2$ (bottom) for GOE; brown curves indicate Poisson statistics. Histograms (top) and points (bottom) correspond to numerical results for $h=0.2$ (a,d), $h=1.0$ (b,e) and $h=3.0$ (c,f).   
    }
    \label{fig:PdsNIAAM}
\end{figure}
In the upper panels we observe that for the non-interacting AA model, the power-spectrum decays as $1/k^2$ for any potential strength. At first glance the results are surprising because one could expect the transition from a metallic phase to an insulating one to be reflected as a change from $1/k$ to $1/k^2$ respectively, however, as it is explicitly explained in our Appendix~\ref{App:A}, the results are related to the non-interacting nature of the model and to the spin-1/2 representation. Once Ising-like interactions between nearest neighbors are activated, the power-spectrum for the interacting AA model (bottom panels) shows a power-law decay resembling pink noise, $1/k$, when the amplitude of the Harper potential is small, $h=0.2$,  [Fig.~\ref{fig:PdsNIAAM}(d)]. When $h=1.2$  [Fig.~\ref{fig:PdsNIAAM}(e)] the system is supposed to be in an intermediate phase  with extended non-ergodic states and one could expect a manifestation of this in the power-spectrum~\cite{Das2022}, but instead, for our system sizes and considered value of $h$, the exponent is already $2$. A deeper analysis of the power-spectrum decay in the intermediate phase is missed and left for future studies. Already at the MBL phase, $h=3.0$ [Fig.~\ref{fig:PdsNIAAM}(f)], the power-law decay is proper of brown noise with decay proportional to $1/k^2$.

In Fig.~\ref{fig:PdsNIAAM} we present complementary results for the level statistics of the non-interacting AA model. We show in the upper panels the level spacing distribution $P(s)$ and the level number variance $\Sigma^2(\ell)$ in the bottom panels. We see that both diagnostics remain Poisson-like for any considered potential strength, $h=0.2$ (a), $h=1.0$ (b) and $h=3.0$ (c). This confirming the scenario already discussed for the power-spectrum of the non-interacting AA model. 
 
\begin{figure}[ht!]
    \centering
    \includegraphics[scale=0.3]{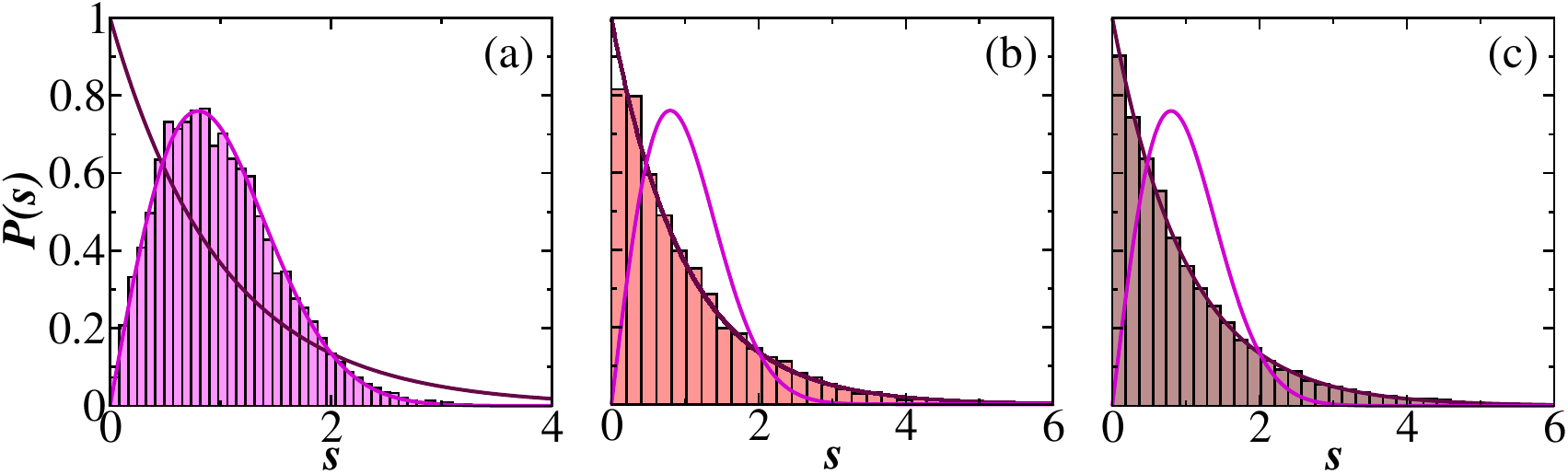}
    \includegraphics[scale=0.3]{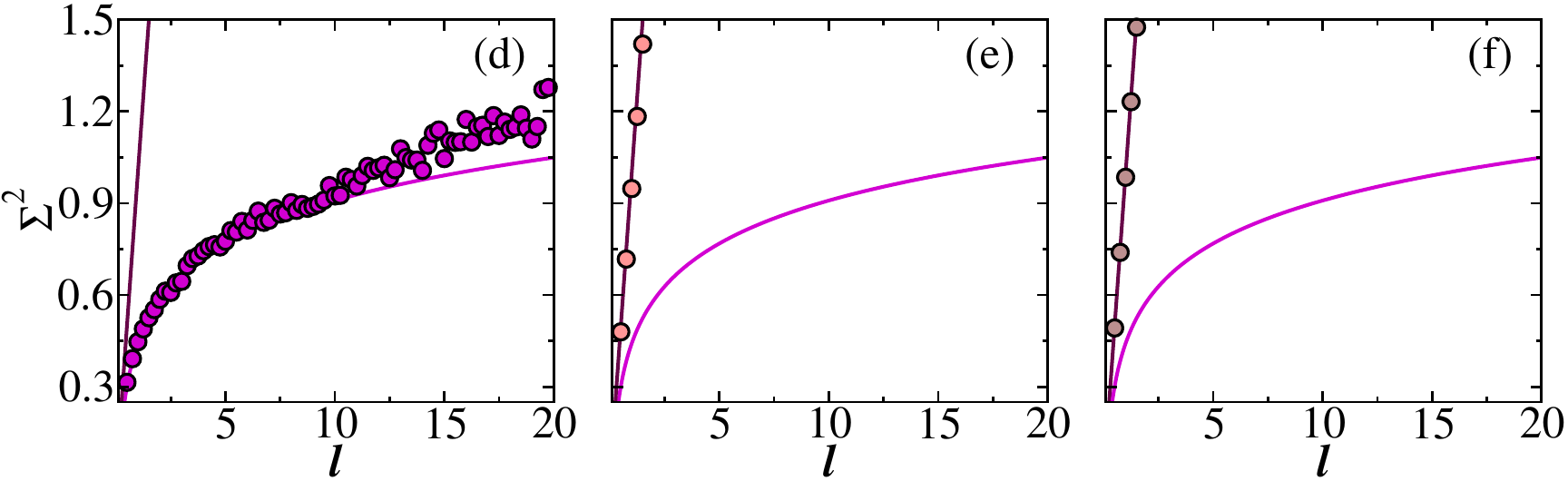}    
    \caption{Distribution of consecutive level spacings (top) and level number variance (bottom) for the interacting AA model. Pink solid curves show Wigner's surmise (top) and $\Sigma^2$ (bottom) for GOE; brown curves indicate Poisson statistics. Histograms (top) and points (bottom) correspond to numerical results for $h=0.2$ (a,d),  $h=1.2$ (b,e) and $h=3.0$ (c,f).  
 }
    \label{fig:PdsIAAM}
\end{figure}

Figure~\ref{fig:PdsIAAM} presents also complementary results but for the interacting AA model. Again, we show in the upper panels the level spacing distribution, while in the bottom panels the level number variance is shown. Both diagnostics display results consistent with the ones obtained for the power-spectrum in Fig.~\ref{fig:PSNIAAM}. For small potential strength, $h=0.2$ [Fig.~\ref{fig:PdsIAAM} (a) and (d)], the level spacing distribution conforms with Wigner's surmise given by Eq.~\eqref{eq:PsWD}, while the level number variance follows the GOE prediction, Eq.~\eqref{eq:LNVGOE}. Certainly, in Fig.~\ref{fig:PdsIAAM} (d) the GOE behavior of $\Sigma^2$ holds only up to the so-called Thouless energy, a well known fact for physical models and in particular for systems with two-body interactions (see for instance~\cite{Schiulaz2019} and references therein). For the next two values of the potential strength, $h=1.2$ and $h=3.0$, in the intermediate and localized phases respectively, the level spacing distribution already looks like Poisson, although a better correspondence is seen for the last value. The level number variance for both cases corresponds also to Poisson-like statistics.

\subsection{Structure and statistics of energy eigenstates}
This section reports results on the structure of energy eigenstates based on the participation ratio, $\text{PR}_\alpha$, through Eq.~\eqref{Eq:PR}. It also presents results for the statistical properties of the energy states, in particular for the distribution of its components. 
%
\begin{figure}[ht!]
    \centering
    \includegraphics[scale=0.3]{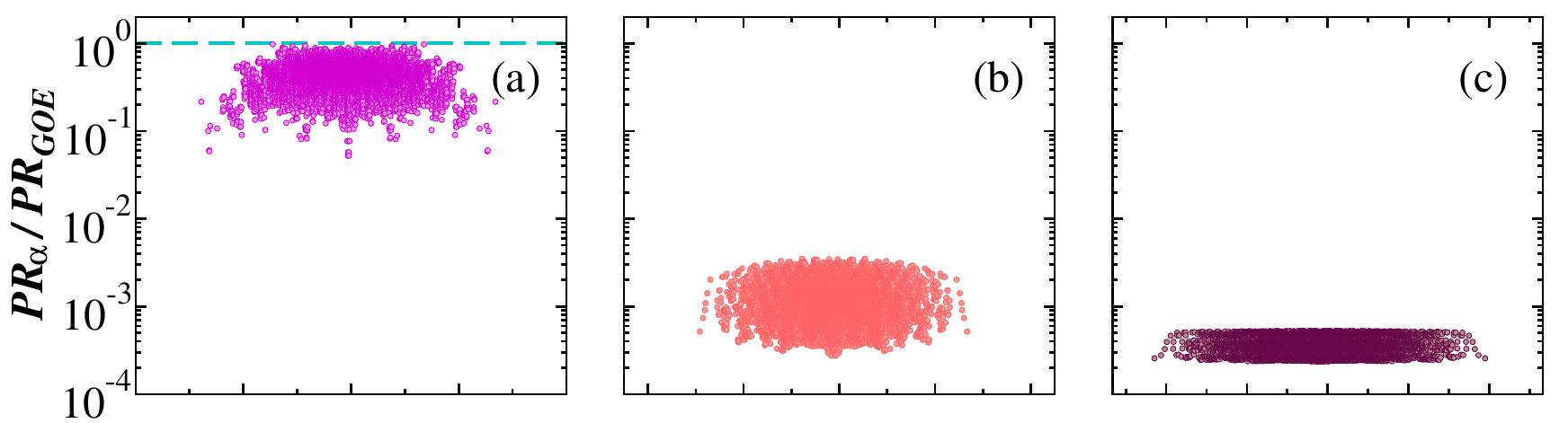}
    \includegraphics[scale=0.3]{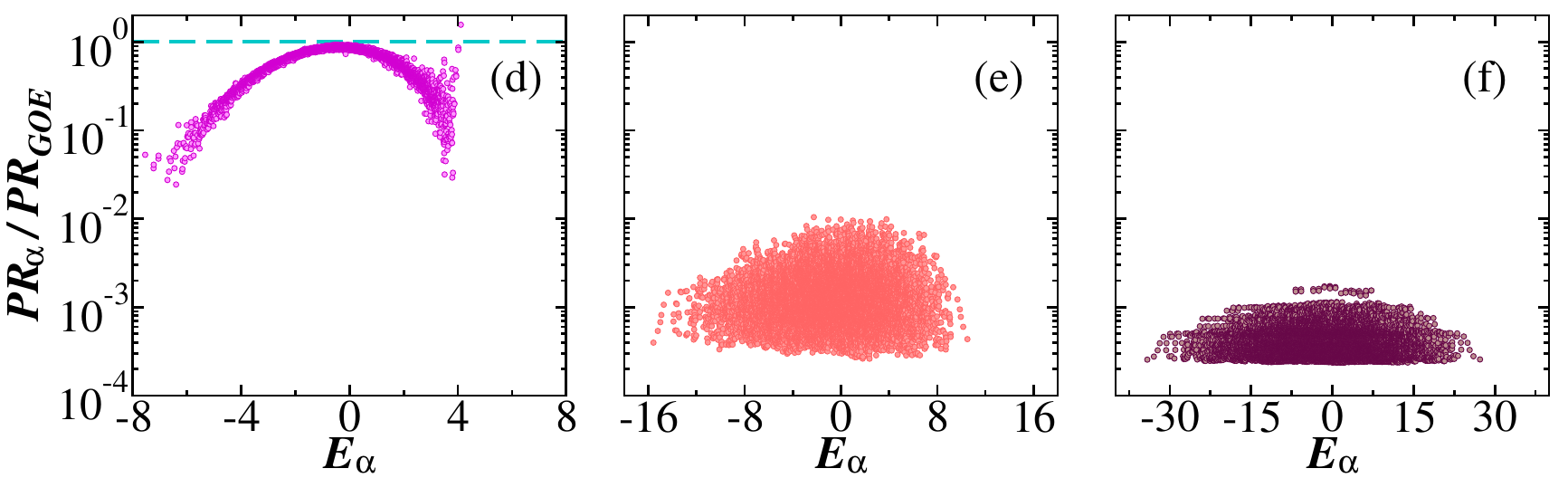}
    \caption{Normalized participation ratio of energy states versus corresponding eigenvalues for the non-interacting (top) and interacting (bottom) AA models. (a) $h=0.2$, (b) $h=1.0$, (c) $h=3.0$, (d) $h=0.2$, (e) $h=1.2$ and (f) $h=3.0$. Dashed lines in (a) and (d) represent  $\text{PR}_\alpha/\text{PR}_\text{GOE}=1$.}
    \label{fig:PRBAAM}
\end{figure}
%
In Fig.~\ref{fig:PRBAAM} we see $\text{PR}_\alpha$ for all energy states, normalized with the theoretical prediction for GOE matrices, $\text{PR}_\text{GOE}$, versus corresponding energy eigenvalues $E_\alpha$. Results are presented for the non-interacting AA model (top panels) and for the interacting AA model (bottom panels). The transition from a metallic-like phase with extended states [Fig.~\ref{fig:PRBAAM} (a) and (d)] to an insulating-like phase with localized states [Fig.~\ref{fig:PRBAAM} (c) and (f)] is evident for both models. In the metallic side of the transition, as it is well known, the states with energy closer to the middle of the spectrum are well extended, just like states of matrices from GOE, while the ones close to the edges of the spectrum are less extended. The $\text{PR}_\alpha$ for the non-interacting model [Fig.~\ref{fig:PRBAAM} (a)] appears highly disperse, this in contrast with the interacting model where the  structure is more compact Fig.~\ref{fig:PRBAAM} (d), specially in a region around the center of the spectrum. This can be considered as an indication that in the extended side only the interacting model will be able to thermalize~\cite{Daleassio2016, Borgonovi2016}. In Fig.~\ref{fig:PRBAAM} (b) and (e) we show results for values of the potential strength at the critical point, $h=1$, for the non-interacting model and for the intermediate phase, $h=1.2$, for the interacting model. In both panels we observe that all energy states are localized with a ratio $\text{PR}_\alpha/\text{PR}_\text{GOE}$ smaller than the unit and broadly distributed. Finally, at $h=3.0$, the states are even more localized with $\text{PR}_\alpha/\text{PR}_\text{GOE}\ll 1$.
\begin{figure}[ht!]
    \centering
    \includegraphics[scale=0.28]{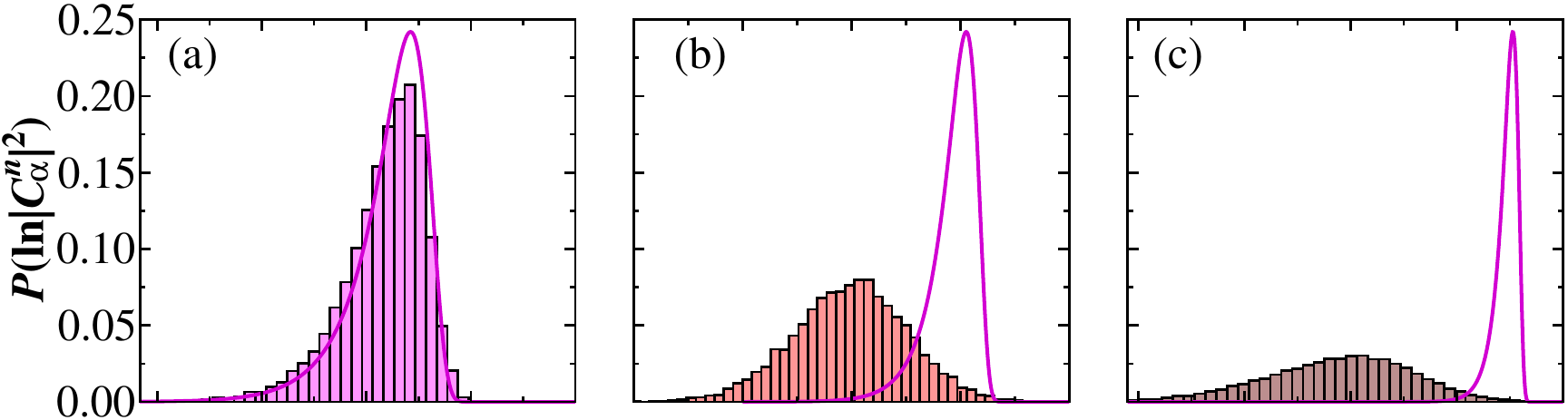}
    \includegraphics[scale=0.28]{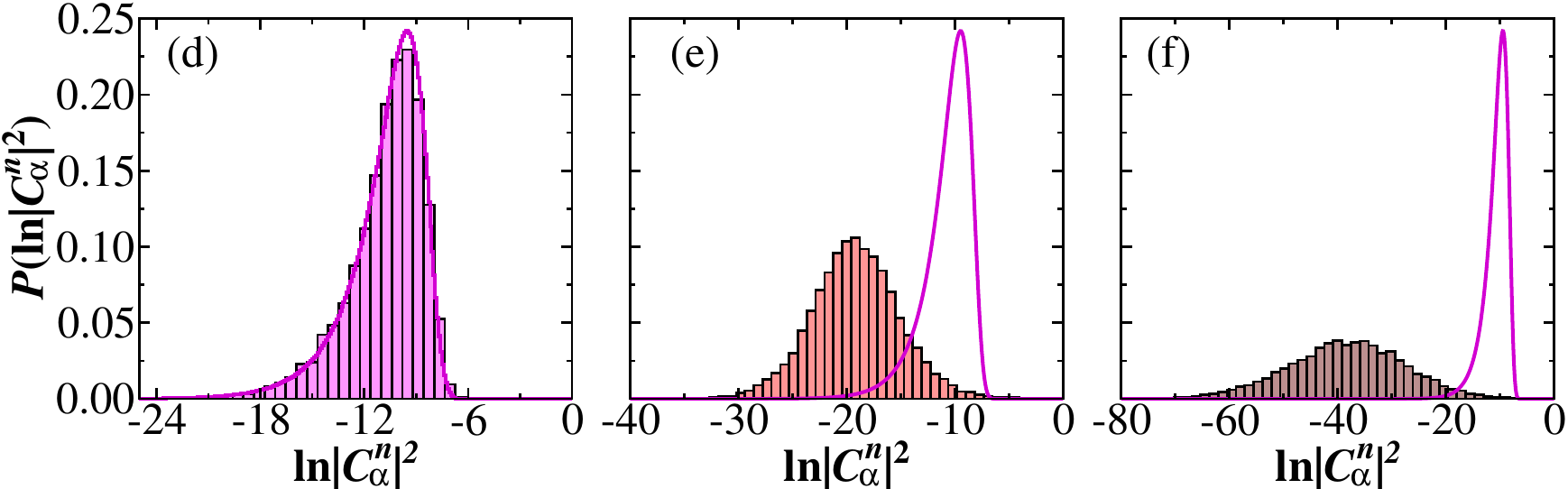}
    \caption{Distribution of $\ln|C_\alpha^n|^2$ for an energy eigenstate with $E_\alpha\approx 0$ of the non-interacting (top) and interacting (bottom) AA models. Solid curves correspond to the Porter-Thomas distribution [Eq.~\eqref{eq:PTD}]. Histograms are for (a) $h=0.2$, (b) $h=1.0$, (c) $h=3.0$, (d) $h=0.2$, (e) $h=1.2$ and (f) $h=3.0$.}
    \label{fig:PTDAAMB}
\end{figure}

We finish this section with the analysis of the distribution of energy states components. The results are presented in Fig.~\ref{fig:PTDAAMB} for the non-interacting AA model (top) and the interacting AA model (bottom). In all panels, the pink solid curve represents the Porter-Thomas distribution given by Eq.~\eqref{eq:PTD}. The histograms are for the components $|C_\alpha^n|^2$ for a state with the energy at the middle of the spectrum, specifically with $E_\alpha\approx 0$. In the extended phase, both models follow the Porter–Thomas distribution [Figs.~\ref{fig:PTDAAMB}(a) and (d)], with better agreement observed for the interacting model. In the intermediate and localized phases, the distribution is non-universal and exhibits an approximately Gaussian shape, indicating that the components are log-normally distributed in these regimes. A Gaussian distribution of $\ln|C_\alpha^n|^2$, corresponding to log-normal statistics, is frequently encountered in disordered systems, including both critical and localized regimes, see for instance Refs.~\cite{Chalker1990,Mirlin2000,Evers2008}).

\section{Conclusions}\label{sec:condisc}
Diagnostics based on level statistics are not able to detect the metal-insulator transition in the non-interacting Aubry-André model when it is written in the language of spin-$1/2$ operators. For any potential strength, no level repulsion is observed and the power-spectrum shows a behavior characteristic of brown noise. The inclusion of nearest-neighbors interactions dramatically changes the image to one with level repulsion at the  metallic side of the transition and absence of repulsion at the localized side. This is also testified by the power-spectrum which corresponds to pink-noise-like in the extended phase and changes to brown-noise-like in the localized phase. In contrast, diagnostics based in energy states are effective to detect the transition in both models, without and with interactions.

\acknowledgments{S.A. M.-C., M. J.-V. and E.J. T.-H. are  grateful to Secihti M{\'e}xico for financial support under project No. CF-2023-I-1748 and to VIEP-BUAP under project No. 00427.}
M. J.-V. and E.J. T.-H. acknowledge useful discussions with Rafael Molina and Adway K. Das.

\appendix
\section{Absence of level repulsion in the spin-$1/2$ non-interacting AAM}
\label{App:A}
\begin{figure}[ht!]
    \centering
    \includegraphics[scale=0.4]{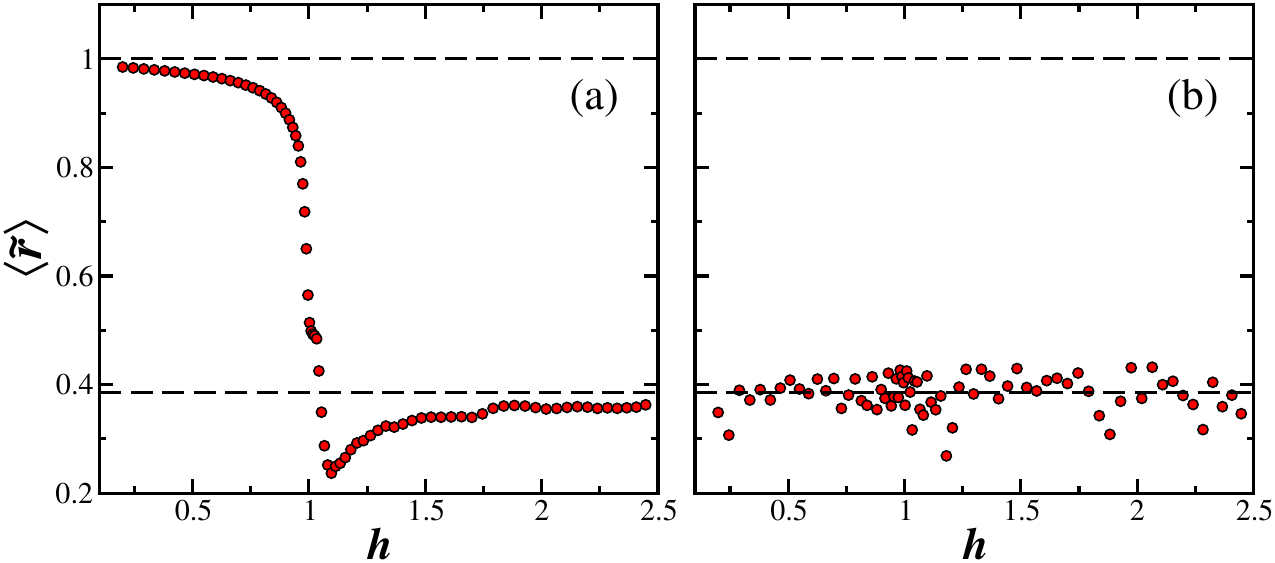}
    \caption{Averaged ratio between consecutive level spacings $\left\langle\tilde{r}\right\rangle$ versus $h$ for the non-interacting Aubry-André model in the single-particle (a) and spin-1/2 (b) representations at half-filling, ${\cal{S}}^z=0$. In both panels, dots are numerical results. Dashed horizontal lines are the theoretical estimates for Poisson-like statistics (bottom) and Picket-fence-like statistics (top). For both models ${\cal{N}}, L = 3432$.}
    \label{fig:rtNIAAM}
\end{figure}

In the main text the absence of repulsion between energy levels at the metallic side of the metal-insulator transition of the non-interacting AA model at half-filling was testified by several diagnostics. One should note that this is not a deficiency of the employed quantities, but instead it is due to the relationship between the many-body spectrum and the single-particle energy levels.
The non-interacting AA model in language of spin-$1/2$ is
\begin{equation}
    H = J \sum_{j=1}^{L-1} (S_j^x S_{j+1}^x + S_j^y S_{j+1}^y) + \sum_{j=1}^{L} h_i S_j^z,
    \label{eq:HamXX}
\end{equation}
where just for simplicity we choose open boundary conditions. Through a Jordan-Wigner transformation applied to Hamiltonian~\eqref{eq:HamXX}, the single-particle Hamiltonian is given by, 
\begin{equation}
    H_\text{SP} = J \sum_{j=1}^{L-1} (c_j^\dagger c_{j+1} + c_{j+1}^\dagger c_j) + \sum_{j=1}^{L} h_j c_j^\dagger c_j,
    \label{eq:Hsp}
\end{equation}
where $c_j^\dagger$ and $c_j$ are creation and annihilation fermionic operator on site $j$. The matrix representation of Hamiltonian~\eqref{eq:Hsp} is tridigonal with elements given by 
\begin{equation}
    \left(H_\text{SP}\right)_{j'j} = h_j\delta_{j'j}+J(\delta_{j'j-1}+\delta_{j'j+1}).
    \label{eq:Hmat}
\end{equation}
In Fig.~\ref{fig:rtNIAAM} we depict the ratio between consecutive level spacings $\left\langle\tilde{r}\right\rangle$~\cite{Oganesyan2007} for the single-particle (a) and spin (b) models.

The difference in the behavior of $\left\langle\tilde{r}\right\rangle$ between both models is clear. For the single-particle model is observed a transition from Picket-fence-like statistics to Poisson-like statistics with values of $\left\langle\tilde{r}\right\rangle=1$ and $\left\langle\tilde{r}\right\rangle=0.386$ respectively~\cite{Atas2013}. A similar scenario was reported for the finite-size one-dimensional Anderson model, see for instance Refs.~\cite{Sorathia2012,Torres2019}. Certainly the correspondence between the prediction for Poisson statistics and the numerical results for large $h$ can be considered not good, but this is only a finite-size effect. In contrast, for the spin model the values of $\left\langle\tilde{r}\right\rangle$ remain fluctuating around the prediction for Poisson statistics, this for any value of $h$.

This last fact is explained because the many-body spectrum $E$ is trivially obtained in terms of the single-particle energy levels $\varepsilon$ as
\begin{equation}\label{eq:summ}
E=\sum_k n_k \varepsilon_k,\quad n_k=0\,\,\text{or}\,\, 1.
\end{equation}
The summation process then erases the existing correlations between single-particle energy levels.
 
We note that it is expected that at the subspace with a single excitation, that is, with ${\cal{S}}^z=(2-L)/2$, the level statistics of the two models should coincide. We verified this, although not shown here.

\bibliography{MSJ_ArXiv}

@article{Abanin2019,
  title={Colloquium: {M}any-body localization, thermalization, and entanglement},
  author={Abanin, Dmitry A and Altman, Ehud and Bloch, Immanuel and Serbyn, Maksym},
  journal={Rev. Mod. Phys},
  volume={91},
  number={2},
  pages={021001},
  year={2019},
  publisher={APS}
}

@article{Altshuler1986,
  title={Repulsion of energy levels and conductivity of small metal samples},
  author={Altshuler, BL and Shklovskii, BI},
  journal={Sov. Phys. JETP},
  volume={64},
  number={1},
  pages={127--135},
  year={1986}
}

@misc{Anderson1958,
  title = {Absence of Diffusion in Certain Random Lattices},
  author = {Anderson, P. W.},
  journal = {Phys. Rev.},
  volume = {109},
  issue = {5},
  pages = {1492--1505},
  year = {1958},
  month = {Mar},
  doi = {10.1103/PhysRev.109.1492}, 
  publisher = {APS}, 
  url ={https://link.aps.org/doi/10.1103/PhysRev.109.1492}
}

@article{Atas2013,
  title={Distribution of the ratio of consecutive level spacings in random matrix ensembles},
  author={Atas, YY and Bogomolny, Eugene and Giraud, O and Roux, G},
  journal={Phys. Rev. Lett.},
  volume={110},
  number={8},
  pages={084101},
  year={2013},
  publisher={APS}
}

@article{Aubry1980,
  title={Analyticity breaking and {A}nderson localization in incommensurate lattices},
  author={Aubry, Serge and Andr{\'e}, Gilles},
  journal={Ann. Israel Phys. Soc},
  volume={3},
  number={133},
  pages={18},
  year={1980}
}

@article{Basko2006,
  title={Metal--insulator transition in a weakly interacting many-electron system with localized single-particle states},
  author={Basko, Denis M and Aleiner, Igor L and Altshuler, Boris L},
  journal={Ann. Phys.},
  volume={321},
  number={5},
  pages={1126--1205},
  year={2006},
  publisher={Elsevier}
}

@article{Berry1977,
  title={Level clustering in the regular spectrum},
  author={Berry, Michael Victor and Tabor, Michael},
  journal={Proc. R. Soc. A: Math. Phys. Eng. Sci.},
  volume={356},
  number={1686},
  pages={375--394},
  year={1977},
  publisher={The Royal Society London}
}

@article{Bertrand2016,
  title={Anomalous {T}houless energy and critical statistics on the metallic side of the many-body localization transition},
  author={Bertrand, Corentin L and Garc{\'\i}a-Garc{\'\i}a, Antonio M},
  journal={Phys. Rev. B},
  volume={94},
  number={14},
  pages={144201},
  year={2016},
  publisher={APS}
}

@article{Bialous2016,
  title={Power spectrum analysis and missing level statistics of microwave graphs with violated time reversal invariance},
  author={Bia{\l}ous, Ma{\l}gorzata and Yunko, Vitalii and Bauch, Szymon and {\L}awniczak, Micha{\l} and Dietz, Barbara and Sirko, Leszek},
  journal={Phys. Rev. Lett.},
  volume={117},
  number={14},
  pages={144101},
  year={2016},
  publisher={APS}
}

@article{Bohigas1984,
  title={Characterization of chaotic quantum spectra and universality of level fluctuation laws},
  author={Bohigas, Oriol and Giannoni, Marie-Joya and Schmit, Charles},
  journal={Phys. Rev. Lett.},
  volume={52},
  number={1},
  pages={1},
  year={1984},
  publisher={APS}
}

@article{Borgonovi2016,
  title={Quantum chaos and thermalization in isolated systems of interacting particles},
  author={Borgonovi, Fausto and Izrailev, Felix M and Santos, Lea F and Zelevinsky, Vladimir G},
  journal={Phys. Rep.},
  volume={626},
  pages={1--58},
  year={2016},
  publisher={Elsevier}
}

@article{Brody1981RMP,
  title   = {Random-Matrix Physics: Spectrum and Strength Fluctuations},
  author  = {Brody, T. A. and Flores, J. and French, J. B. and Mello, P. A. and Pandey, A. and Wong, S. S. M.},
  journal = {Rev. Mod. Phys.},
  volume  = {53},
  pages   = {385--479},
  year    = {1981},
  doi     = {10.1103/RevModPhys.53.385}
}

@article{Casati1980,
  title={On the connection between quantization of nonintegrable systems and statistical theory of spectra},
  author={Casati, G and Valz-Gris, F and Guarnieri, I},
  journal={Lett. Nuovo Cimento (1971-1985)},
  volume={28},
  pages={279--282},
  year={1980},
  publisher={Societ{\`a} Italiana di Fisica}
}

@article{Casati1985,
  title={Energy-level statistics of integrable quantum systems},
  author={Casati, Giulio and Chirikov, BV and Guarneri, Italo},
  journal={Physical review letters},
  volume={54},
  number={13},
  pages={1350},
  year={1985},
  publisher={APS}
}

@article{Chalker1990,
  title={Scaling and eigenfunction correlations near a mobility edge},
  author={Chalker, John T},
  journal={Physica A: Stat. Mech. Appl.},
  volume={167},
  number={1},
  pages={253--258},
  year={1990},
  publisher={Elsevier}
}

@article{Corps2020,
  title={Thouless energy challenges thermalization on the ergodic side of the Many-body localization transition},
  author={Corps, {\'A}ngel L and Molina, Rafael Armando and Rela{\~n}o, Armando},
  journal={Phys. Rev. B},
  volume={102},
  number={1},
  pages={014201},
  year={2020},
  publisher={APS}
}

@article{Corps2021,
  title={Signatures of a critical point in the Many-body localization transition},
  author={Corps, {\'A}ngel L and Molina, Rafael Armando and Rela{\~n}o, Armando},
  journal={SciPost Phys.},
  volume={10},
  number={5},
  pages={107},
  year={2021}
}

@article{Daleassio2016,
  title={From quantum chaos and eigenstate thermalization to statistical mechanics and thermodynamics},
  author={D'Alessio, Luca and Kafri, Yariv and Polkovnikov, Anatoli and Rigol, Marcos},
  journal={Adv. Phys.},
  volume={65},
  number={3},
  pages={239--362},
  year={2016},
  publisher={Taylor \& Francis}
}

@article{Das2022,
  title={Nonergodic extended states in the $\beta$ ensemble},
  author={Das, Adway Kumar and Ghosh, Anandamohan},
  journal={Phys. Rev. E},
  volume={105},
  number={5},
  pages={054121},
  year={2022},
  publisher={APS}
}

@article{Evers2008,
  title={Anderson transitions},
  author={Evers, Ferdinand and Mirlin, Alexander D},
  journal={Reviews of Modern Physics},
  volume={80},
  number={4},
  pages={1355--1417},
  year={2008},
  publisher={APS}
}

@article{Faleiro2004,
  title={Theoretical derivation of $1/f$ noise in Quantum Chaos},
  author={Faleiro, E and G{\'o}mez, JMG and Molina, RA and Mu{\~n}oz, L and Rela{\~n}o, A and Retamosa, J},
  journal={Phys. Rev. Lett.},
  volume={93},
  number={24},
  pages={244101},
  year={2004},
  publisher={APS}
}

@article{Faleiro2006,
  title={Power spectrum analysis of experimental {S}inai quantum billiards},
  author={Faleiro, E and Kuhl, U and Molina, RA and Mu{\~n}oz, L and Rela{\~n}o, A and Retamosa, J},
  journal={Phys. Lett. A},
  volume={358},
  number={4},
  pages={251--255},
  year={2006},
  publisher={Elsevier}
}

@article{Feingold1985,
  title={Energy-level statistics of integrable quantum systems},
  author={Feingold, Mario},
  journal={Phys. Rev. Lett.},
  volume={55},
  number={23},
  pages={2626},
  year={1985},
  publisher={APS}
}

@article{Gomez2005,
  title={$1/f^{\alpha}$ noise in spectral fluctuations of quantum systems},
  author={G{\'o}mez, JMG and Rela{\~n}o, A and Retamosa, J and Faleiro, E and Salasnich, L and Vrani{\v{c}}ar, M and Robnik, M},
  journal={Phys. Rev. Lett.},
  volume={94},
  number={8},
  pages={084101},
  year={2005},
  publisher={APS}
}

@article{Gubin2012,
  title={Quantum chaos: An introduction via chains of interacting spins 1/2},
  author={Gubin, Aviva and F Santos, Lea},
  journal={American Journal of Physics},
  volume={80},
  number={3},
  pages={246--251},
  year={2012},
  publisher={AIP Publishing}
}

@article{Guhr1998,
  title={Random-matrix theories in quantum physics: common concepts},
  author={Guhr, Thomas and M{\"u}ller--Groeling, Axel and Weidenm{\"u}ller, Hans A},
  journal={Phys. Rep.},
  volume={299},
  number={4-6},
  pages={189--425},
  year={1998},
  publisher={Elsevier}
}

@article{Harper1955,
  title={Single band motion of conduction electrons in a uniform magnetic field},
  author={Harper, Philip George},
  journal={Proc. Phys. Soc. Section A},
  volume={68},
  number={10},
  pages={874},
  year={1955},
  publisher={IOP Publishing}
}

@article{Hofstetter1993,
  title={Statistical properties of the eigenvalue spectrum of the three-dimensional {A}nderson Hamiltonian},
  author={Hofstetter, E and Schreiber, M},
  journal={Phys. Rev. B},
  volume={48},
  number={23},
  pages={16979},
  year={1993},
  publisher={APS}
}

@article{Kuhl1998,
  title={Microwave realization of the {H}ofstadter butterfly},
  author={Kuhl, U and St{\"o}ckmann, H-J},
  journal={Phys. Rev. Lett.},
  volume={80},
  number={15},
  pages={3232},
  year={1998},
  publisher={APS}
}

@article{Li2023,
  title={Observation of critical phase transition in a generalized {A}ubry-{A}ndr{\'e}-{H}arper model with superconducting circuits},
  author={Li, Hao and Wang, Yong-Yi and Shi, Yun-Hao and Huang, Kaixuan and Song, Xiaohui and Liang, Gui-Han and Mei, Zheng-Yang and Zhou, Bozhen and Zhang, He and Zhang, Jia-Chi and others},
  journal={npj Quantum Inf.},
  volume={9},
  number={1},
  pages={40},
  year={2023},
  publisher={Nature Publishing Group UK London}
}

@article{Luitz2015,
  title={Many-body localization edge in the random-field {H}eisenberg chain},
  author={Luitz, David J and Laflorencie, Nicolas and Alet, Fabien},
  journal={Phys. Rev. B},
  volume={91},
  number={8},
  pages={081103},
  year={2015},
  publisher={APS}
}

@article{McDonald1979,
  title={Spectrum and eigenfunctions for a Hamiltonian with stochastic trajectories},
  author={McDonald, Steven W and Kaufman, Allan N},
  journal={Phys. Rev. Lett.},
  volume={42},
  number={18},
  pages={1189},
  year={1979},
  publisher={APS}
}

@article{Mirlin2000,
  title={Multifractality and critical fluctuations at the {A}nderson transition},
  author={Mirlin, AD and Evers, Ferdinand},
  journal={Phys. Rev. B},
  volume={62},
  number={12},
  pages={7920},
  year={2000},
  publisher={APS}
}

@article{Mott1968,
  title={Metal-insulator transition},
  author={Mott, Nevill F},
  journal={Reviews of Modern Physics},
  volume={40},
  number={4},
  pages={677},
  year={1968},
  publisher={APS}
}

@article{Nandkishore2015,
  title={Many-body localization and thermalization in quantum statistical mechanics},
  author={Nandkishore, Rahul and Huse, David A},
  journal={Annu. Rev. Condens. Matter Phys.},
  volume={6},
  number={1},
  pages={15--38},
  year={2015},
  publisher={Annual Reviews}
}

@article{Oganesyan2007,
  title={Localization of interacting fermions at high temperature},
  author={Oganesyan, Vadim and Huse, David A},
  journal={Phys. Rev. B},
  volume={75},
  number={15},
  pages={155111},
  year={2007},
  publisher={APS}
}

@article{Pachon2018,
  title={Origin of the $1/f^\alpha$ spectral noise in chaotic and regular quantum systems},
  author={Pach{\'o}n, Leonardo A and Rela{\~n}o, Armando and Peropadre, Borja and Aspuru-Guzik, Al{\'a}n},
  journal={Phys. Rev. E},
  volume={98},
  number={4},
  pages={042213},
  year={2018},
  publisher={APS}
}

@article{Paladino2014,
  title={$1/f$ noise: Implications for solid-state quantum information},
  author={Paladino, E and Galperin, YM and Falci, G and Altshuler, BL},
  journal={Rev. Mod. Phys.},
  volume={86},
  number={2},
  pages={361--418},
  year={2014},
  publisher={APS}
}

@article{Porter1956,
  title = {Fluctuations of Nuclear Reaction Widths},
  author = {Porter, C. E. and Thomas, R. G.},
  journal = {Phys. Rev.},
  volume = {104},
  issue = {2},
  pages = {483--491},
  numpages = {0},
  year = {1956},
  month = {Oct},
  publisher = {American Physical Society},
  doi = {10.1103/PhysRev.104.483},
  url = {https://link.aps.org/doi/10.1103/PhysRev.104.483}
}

@article{Press1978,
  title={Flicker noises in astronomy and elsewhere},
  author={Press, William H},
  journal={Comments on Modern Physics, Part C-Comments on Astrophysics, vol. 7, no. 4, 1978, p. 103-119.},
  volume={7},
  pages={103--119},
  year={1978}
}

@article{Relano2002,
  title={Quantum chaos and $1/f$ noise},
  author={Relano, A and G{\'o}mez, JMG and Molina, RA and Retamosa, J and Faleiro, E},
  journal={Phys. Rev. Lett.},
  volume={89},
  number={24},
  pages={244102},
  year={2002},
  publisher={APS}
}

@article{Relano2008,
  title={Power-spectrum characterization of the continuous {G}aussian ensemble},
  author={Rela{\~n}o, Armando and Mu{\~n}oz, L and Retamosa, J and Faleiro, E and Molina, Rafael A},
  journal={Phys. Rev. E},
  volume={77},
  number={3},
  pages={031103},
  year={2008},
  publisher={APS}
}

@article{Roati2008,
  title={{A}nderson localization of a non-interacting {B}ose--{E}instein condensate},
  author={Roati, Giacomo and D’Errico, Chiara and Fallani, Leonardo and Fattori, Marco and Fort, Chiara and Zaccanti, Matteo and Modugno, Giovanni and Modugno, Michele and Inguscio, Massimo},
  journal={Nature},
  volume={453},
  number={7197},
  pages={895--898},
  year={2008},
  publisher={Nature Publishing Group UK London}
}

@article{Thouless1977,
  title={Maximum metallic resistance in thin wires},
  author={Thouless, DJ},
  journal={Phys. Rev. Lett.},
  volume={39},
  number={18},
  pages={1167},
  year={1977},
  publisher={APS}
}

@article{Torres2015,
  title={Dynamics at the many-body localization transition},
  author={Torres-Herrera, EJ and Santos, Lea F},
  journal={Phys. Rev. B},
  volume={92},
  number={1},
  pages={014208},
  year={2015},
  publisher={APS}
}

@article{Santos2009,
  title={Transport and control in one-dimensional systems},
  author={Santos, Lea F},
  journal={J. Math. Phys.},
  volume={50},
  number={9},
  year={2009},
  publisher={AIP Publishing}
}

@article{Scaramazza2016,
  title={Integrable matrix theory: Level statistics},
  author={Scaramazza, Jasen A and Shastry, B Sriram and Yuzbashyan, Emil A},
  journal={Phys. Rev. E},
  volume={94},
  number={3},
  pages={032106},
  year={2016},
  publisher={APS}
}

@article{Seligman1986,
  title={Energy-Level Statistics of Integrable Quantum Systems},
  author={Seligman, TH and Verbaarschot, JJM},
  journal={Phys. Rev. Lett.},
  volume={56},
  number={25},
  pages={2767},
  year={1986},
  publisher={APS}
}

@article{Schiulaz2019,
  title={Thouless and relaxation time scales in many-body quantum systems},
  author={Schiulaz, Mauro and Torres-Herrera, E Jonathan and Santos, Lea F},
  journal={Phys. Rev. B},
  volume={99},
  number={17},
  pages={174313},
  year={2019},
  publisher={APS}
}

@article{Serbyn2016,
  title={Spectral statistics across the many-body localization transition},
  author={Serbyn, Maksym and Moore, Joel E},
  journal={Phys. Rev. B},
  volume={93},
  number={4},
  pages={041424},
  year={2016},
  publisher={APS}
}

@article{Sorathia2012,
  title={From closed to open one-dimensional {A}nderson model: Transport versus spectral statistics},
  author={Sorathia, S and Izrailev, FM and Zelevinsky, VG and Celardo, GL},
  journal={Phys. Rev. E},
  volume={86},
  number={1},
  pages={011142},
  year={2012},
  publisher={APS}
}

@article{Sorensen1991,
  title={Level-spacing statistics for the {A}nderson model in one and two dimensions},
  author={S{\"o}rensen, Mads Peter and Schneider, Toni},
  journal={Z. Phys. B: Condens. Matter},
  volume={82},
  number={1},
  pages={115--119},
  year={1991},
  publisher={Springer}
}

@article{Torres2017ptrs,
  title={Dynamical manifestations of quantum chaos: correlation hole and bulge},
  author={Torres-Herrera, EJ and Santos, Lea F},
  journal={Phil. Trans. R. Soc. A, Phil. Trans. Math. Phys. Eng. Sci.},
  volume={375},
  number={2108},
  pages={20160434},
  year={2017},
  publisher={The Royal Society Publishing}
}

@article{Torres2017,
author = {Torres-Herrera, E. J. and Santos, Lea F.},
title = {Extended nonergodic states in disordered many-body quantum systems},
journal = {Ann. Phys},
volume = {529},
number = {7},
pages = {1600284},
year = {2017}
}

@article{Torres2018,
  title={Generic dynamical features of quenched interacting quantum systems: Survival probability, density imbalance, and out-of-time-ordered correlator},
  author={Torres-Herrera, EJ and Garc{\'\i}a-Garc{\'\i}a, Antonio M and Santos, Lea F},
  journal={Phys. Rev. B},
  volume={97},
  number={6},
  pages={060303},
  year={2018},
  publisher={APS}
}

@article{Torres2019,
  title={Level repulsion and dynamics in the finite one-dimensional {A}nderson model},
  author={Torres-Herrera, E Jonathan and M{\'e}ndez-Berm{\'u}dez, JA and Santos, Lea F},
  journal={Phys. Rev. E},
  volume={100},
  number={2},
  pages={022142},
  year={2019},
  publisher={APS}
}

@article{Torres2020,
  title={Dynamical detection of level repulsion in the one-particle {A}ubry-{A}ndr{\'e} model},
  author={Torres-Herrera, Eduardo Jonathan and Santos, Lea F},
  journal={Condensed Matter},
  volume={5},
  number={1},
  pages={7},
  year={2020},
  publisher={MDPI}
}

@article{Ulla196465,
title = {Invariance hypothesis and higher correlations of Hamiltonian matrix elements},
journal = {Nucl. Phys.},
volume = {58},
pages = {65-71},
year = {1964},
issn = {0029-5582},
author = {Nazakat Ullah}
}

@article{Voss1978,
  title={``$1/f$ noise'' in music: Music from $1/f$ noise},
  author={Voss, Richard F and Clarke, John},
  journal={J. Acoust. Soc. Am.},
  volume={63},
  number={1},
  pages={258--263},
  year={1978},
  publisher={Acoustical Society of America}
}

@article{Vsuntajs2020,
  title={Quantum chaos challenges many-body localization},
  author={{\v{S}}untajs, Jan and Bon{\v{c}}a, Janez and Prosen, Toma{\v{z}} and Vidmar, Lev},
  journal={Phys. Rev. E},
  volume={102},
  number={6},
  pages={062144},
  year={2020},
  publisher={APS}
}

@article{Weissman1988,
  title={$1/f$ noise and other slow, nonexponential kinetics in condensed matter},
  author={Weissman, MB},
  journal={Rev. Mod. Phys},
  volume={60},
  number={2},
  pages={537},
  year={1988},
  publisher={APS}
}

@article{West1990,
  title={The noise in natural phenomena},
  author={West, Bruce J and Shlesinger, Michael},
  journal={Am. Sci.},
  volume={78},
  number={1},
  pages={40--45},
  year={1990},
  publisher={JSTOR}
}

@article{Xu2019,
  title = {Butterfly effect in interacting {A}ubry-{A}ndre model: Thermalization, slow scrambling, and many-body localization},
  author = {Xu, Shenglong and Li, Xiao and Hsu, Yi-Ting and Swingle, Brian and Das Sarma, S.},
  journal = {Phys. Rev. Res.},
  volume = {1},
  issue = {3},
  pages = {032039},
  numpages = {6},
  year = {2019},
  month = {Dec},
  publisher = {American Physical Society},
  doi = {10.1103/PhysRevResearch.1.032039},
  url = {https://link.aps.org/doi/10.1103/PhysRevResearch.1.032039}
}

@article{Yang2024,
  title={Level-spacing distribution of localized phases induced by quasiperiodic potentials},
  author={Yang, Chao and Wang, Yucheng},
  journal={Physical Review B},
  volume={109},
  number={21},
  pages={214210},
  year={2024},
  publisher={APS}
}

@article{Zelevinsky1996,
  title   = {Quantum Chaos and Complexity in Nuclei},
  author  = {Zelevinsky, Vladimir G.},
  journal = {Annu. Rev. Nucl. Part. Sci.},
  volume  = {46},
  pages   = {237--279},
  year    = {1996},
}

@article{Talib2025,
  title={Spectral form factor and power spectrum for trapped rotating interacting bosons: An exact diagonalization study},
  author={Talib, Mohd and Ahsan, MAH},
  journal={arXiv preprint arXiv:2512.00828},
  year={2025}
}

@book{Abrahams2010,
  author = {Abrahams, Elihu},
  title = {50 Years of {A}nderson Localization},
  publisher = {WORLD SCIENTIFIC},
  year = {2010},
  doi = {10.1142/7663},
  address = {},
  edition   = {},
  URL = {https://www.worldscientific.com/doi/abs/10.1142/7663},
  eprint = {https://www.worldscientific.com/doi/pdf/10.1142/7663}
}

@book{Mehta2004,
  title={Random matrices},
  author={M. L. Mehta},
  year={2004},
  publisher={Elsevier}
}

@book{Porter1965,
  title     = {Statistical Theories of Spectra: Fluctuations},
  editor    = {Porter, Charles E.},
  series    = {Perspectives in Physics},
  publisher = {Academic Press},
  address   = {New York},
  year      = {1965},
  pages     = {xv+576},
  isbn      = {9780125623568}
}

\end{document}